\documentclass[12pt]{article}
\usepackage{graphicx}  
\usepackage{pdfpages}
\usepackage{amsmath}   
\usepackage{amssymb}   
\usepackage{bm} 
\usepackage{dcolumn}
\usepackage{color}
\usepackage{mathrsfs}
\usepackage{amsfonts}
\usepackage{varioref}
\usepackage[utf8]{inputenc}
\usepackage{amsmath}
\usepackage{amsfonts}
\usepackage{amssymb}
\usepackage{enumitem}
\usepackage{float}
\usepackage{epsfig}
\usepackage{amsfonts}
\usepackage{latexsym}
\usepackage{hyperref}
\usepackage{setspace}
\usepackage{bm}
\usepackage{slashed}
\usepackage{graphicx}
\usepackage{bm}
\usepackage{hyperref}
\usepackage{dsfont}
\usepackage{caption}
\usepackage{subcaption}

\title{\bf Optomechanical non-reciprocity and its equivalence to antiresonance: Control of isolation frequency using mechanical drive }
\author{Chetan Waghela \footnote{chetan.waghela@iitrpr.ac.in}, ~~Shubhrangshu Dasgupta \\
{{\small  Department of Physics, Indian Institute of Technology Ropar}}\\ {{\small Rupnagar, Punjab 140 001,
India}}}           

\date{\today}

\begin{document}

\maketitle

\begin{abstract}
We demonstrate that optomechanical non-reciprocity is equivalent to the anti-resonance, often discussed in the context of coupled driven harmonic oscillators. We show that that suitable phase-difference between the cavity driving fields make the relevant optomechanical couplings complex, which leads to non-reciprocity in the field fluctuations and anti-resonance in average field amplitudes. This analogy with anti-resonance demonstrates that only for a particular frequency (the so-called isolation frequency) of input signal, maximum non-reciprocity can be achieved. In contrast to the previous studies, we here show that one can dynamically control this isolation frequency by applying a mechanical drive of suitable frequency to the membrane in the optomechanical setup.
\end{abstract}

\section{Introduction}
An atomic system behaves as an opaque medium for an input light field, at resonance. In presence of suitable control field, this system may also appear transparent at the same frequency. Such quantum optical phenomenon, often referred to as the electromagnetically induced transparency (EIT), can be explained in terms of quantum interference between relevant transition amplitudes. Importantly, EIT refers to transparency for both the directions of the field, i.e., positive and negative directions of the quantization axis. 

On the other hand, an optical isolator can be treated as an optical analogue of an electronic diode. As the diodes allow the electrons to move predominantly along one direction and not in the opposite direction, an optical isolator lets the photons transmit in only one direction. This means, unlike EIT, the system becomes transparent for only one direction of the field. Quantitatively, the input and output mode of an ideal two-port optical isolator can be related via a scattering matrix $S$, as
\begin{equation} 
\Bar{A}_{out}=S \Bar{A}_{in}\\ \text{ where }
   S = \begin{pmatrix}
 0 & 1 \\
 0 & 0 \label{eq1}
\end{pmatrix} \text{or} 
 \begin{pmatrix}
 0 & 0 \\
 1 & 0 
\end{pmatrix}
\end{equation}
where, $\Bar{A}_{in}$ and $\Bar{A}_{out}$ are the column vectors representing the input and output modes of the device.

However, it is challenging to construct an isolator for light (or photons) due to the Lorentz reciprocity theorem. According to this theorem, a device with linear, isotropic and time-independent dielectric constant \cite{Jalas-2013,Caloz-2018} cannot be non-reciprocal and hence cannot be used for isolation.
In addition, an ideal optical diode (a) should function for any arbitrary frequency of light, i.e., the matrix $S$ should be frequency-independent, (b)  does not require an external control field or a bias (note that the so-called Faraday isolators, on the other hand, require an external magnetic field to operate \cite{book:75551}), (c) should suffer no loss in the direction in which the device is transparent. 

\begin{center}
\begin{figure}
\begin{center}
 \includegraphics[width=.6\textwidth]{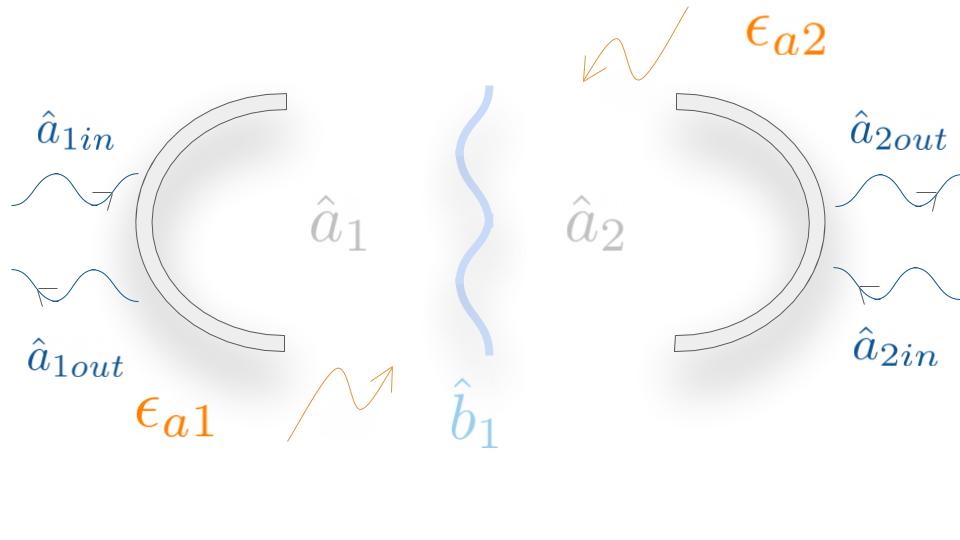} \\
 \caption{Generic setup for an optomechanical isolator.}
 \label{Fig1}
 \end{center}
 \end{figure}
\end{center}

In recent years, optomechanical isolators have emerged as possible candidates for achieving isolation \cite{Xu-2015,Verhagen-2017,Habraken-2012,Metelmann-2015,Miri-2017}. Consider that two optical modes are coupled to a single mechanical oscillator, while the optical modes are driven by two control fields [see Fig. \ref{Fig1}]. An input field from the optical mode 1 to 2 via the mechanical oscillator will acquire an overall phase equal to the phase difference between the control fields, whereas in the opposite direction it acquires an opposite phase. This directional phase difference is equivalent to the Peierls phase \cite{Hofstadter-1976} for charged particles in a (magnetic) gauge potential, and it can be interpreted as a synthetic magnetic flux biasing the system \cite{Verhagen-2017,Fang-2017}. It is seen that in these systems for particular input signal frequencies and particular phase difference between the control fields, the scattering matrix of the isolator is as given by Eq. $\eqref{eq1}$.

This has been further shown \cite{Hemmer-1988, Alzar-2002, Souza-2015} that EIT exhibits a similar intensity spectrum, as found in case of anti-resonance in  classical systems, e.g., the coupled pendulum. \cite{book:2130260}. In fact, mechanical analogue of Fano resonances and some other optical phenomena have been thoroughly investigated in \cite{Satpathy-2012,Novotny-2010,book:289457, foulaadvand2010mechanical}. In this paper, we ask the following question: can we interpret the optical isolation, as well, in terms of anti-resonance ? We analyze a generic optoemechanical system \cite{Xu-2015} and find the answer to this question as affirmative. Our result further emphasizes that both the reciprocal phenomenon like EIT as well as the optical non-reciprocal effects, namely, optical isolation, have the same origin - the anti-resonance. 


\begin{center}
\begin{figure*}
\begin{center}
 \includegraphics[width=.8\textwidth]{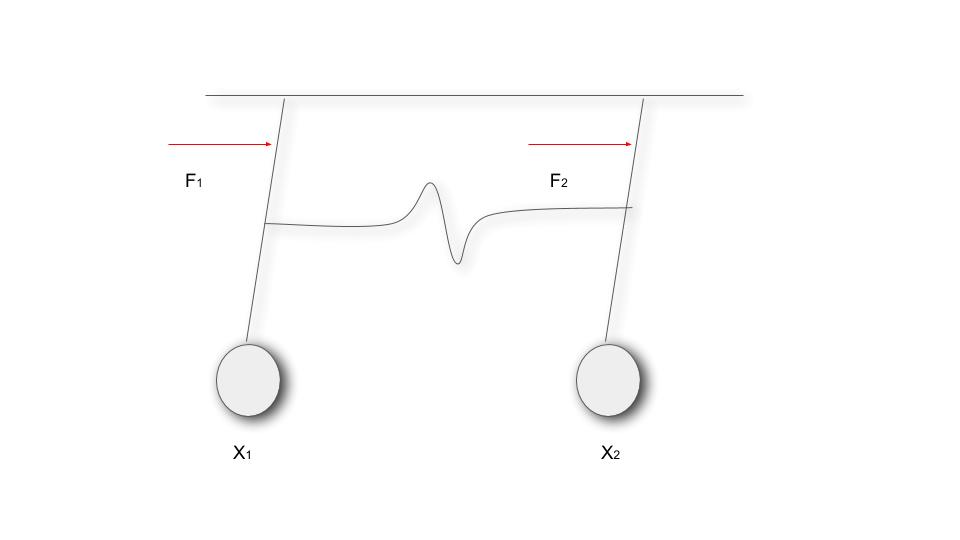} \\
 \caption{A schematic diagram of a coupled pendulum with driving forces $F_1\cos(\omega_{d1} t + \phi_1)$ and $F_2\cos(\omega_{d2} t + \phi_2)$.}
 \label{Fig2}
 \end{center}
 \end{figure*}
\end{center}

Though the nonreciprocal devices behave in the principle of anti-resonance, they exhibit maximum isolation only at a particular input signal frequency (called `isolation frequency'). Thus the broadband feature, as required for an ideal isolator, is compromised \cite{Xu-2015,Verhagen-2017,Habraken-2012,Miri-2017}. In fact, in these works, isolation is observed to be dependent on the natural frequency of the mechanical resonator. This natural frequency is often fixed by the fabrication and material properties of the resonator and thus is not externally controllable. In this paper, we demonstrate that it is possible to control the isolation frequency, in a dynamical fashion, by a mechanical drive connected to the setup. 

The structure of this paper is as follows: In Sec. \ref{Section2}, we analyze a mechanical system and demonstrate the phase-dependent anti-resonance. In Sec. \ref{Section3}, we consider an optomechanical system with a mechanical driving. We explicitly show how the non-reciprocity is related to the inherent anti-resonance in the system. We further explore how one can control the isolation frequency of the device. We conclude the paper in Sec. \ref{Section4},

\section{\label{Section2}Anti-resonance in coupled pendulum}


Usually, a pair of simple pendulums (each of angular frequency $\omega$), linearly coupled with the corresponding constant $g$ can have two independent eigenmodes of oscillation - the in-phase mode at a frequency $\omega$ and the out-of-phase mode at a frequency $\sqrt{\omega^2+g^2}$. The intensity spectrum of such a system therefore exhibits a doublet, with a peak separation $\sim g$ in weak coupling limit. 

Anti-resonance occurs when one of these pendulums is driven by an oscillating field. The amplitude spectrum of the driven oscillator exhibits a deep minimum at a frequency half-way between the two normal-mode frequencies. More importantly, this is associated with a sharp change in phase across this anti-resonance point.

Here we consider both the pendulums to be driven with oscillating fields [see Fig. \ref{Fig2}]. Generalizing to the case, when the pendulums are of different natural frequencies $\omega_1$ and $\omega_2$, we can write down the corresponding equations of motion of the amplitudes $x_i$ of these pendulums, as follows:
\begin{equation}
 \begin{split}
     \ddot{x}_1+2 \gamma_1 \dot{x}_1 - 2  g \omega_1 x_2 + \omega_1^2 x_1 = 2 F_1 \cos(\omega_{d,1} t + \phi_1)\;, \\
     \ddot{x}_2+2 \gamma_2 \dot{x}_2 - 2 g \omega_2 x_1 + \omega_2^2 x_2 = 2 F_2 \cos(\omega_{d,2} t + \phi_2)\;,
     \label{eq2}
     \end{split}
     \end{equation}
where, $\gamma_i$ denotes the damping rate of the $i$th pendulum $(i\in 1,2)$, the $g$ denotes the coupling constant between the two pendulums, and $F_i$ denotes the amplitude of the $i$th driving force with frequency $\omega_{d,i}$ and phase $\phi_i$.

These equations (\ref{eq2}) involve second-order derivatives of $x_i$'s. Next, we choose a suitable change of variables as $\alpha_1=\omega_1 x_1 + i \dot{x_1}$ and $\alpha_2=\omega_2 x_2 + i \dot{x_2}$. Assuming that the driving fields have the same frequencies, i.e., $\omega_{d,1}=\omega_{d,2}=\omega_{d}$ and making a transformation to the rotating frame with respect to the driving field frequency $\omega_d$, i.e.,  $\alpha_i \rightarrow \alpha_i e ^{-i \omega_{d} t}$,
 $\dot{\alpha}_i \rightarrow \dot{\alpha}_i e ^{-i \omega_{d} t}-i \omega_{d} \alpha_i e^{i \omega_{d} t }$ and $\Delta_i=\omega_i-\omega_{d}$ 
we get
\begin{equation}
 \begin{split}
 &\dot{\alpha}_1=-i \Delta_1 \alpha_1 - \gamma_1 (\alpha_1-\alpha_1^* e^{2 i \omega_{d} t }) - i  \frac{g\omega_1}{\omega_2}(\alpha_2 + \alpha_2^* e^{2 i \omega_{d} t} \\
 &+ i F_1 (e^{i \phi_1} + e^{-2 i \omega_{d} t} e^{-i \phi_1})\;. \\
 &\dot{\alpha}_2=-i \Delta_2 \alpha_2 - \gamma_2 (\alpha_2-\alpha_2^* e^{2 i \omega_{d} t }) - i  \frac{g\omega_2}{\omega_1}(\alpha_1 + \alpha_1^* e^{2 i \omega_{d} t} \\
 &+ i F_2 (e^{i \phi_2} + e^{-2 i \omega_{d} t} e^{-i \phi_2})\;.
 \label{eq3}
 \end{split}
\end{equation}


Neglecting the counter-rotating terms containing $e^{2 i \omega_{d} t}$ (which corresponds to using the rotating wave approximation), we get the final set of equations, as follows:
\begin{equation}
 \begin{split}
\dot{\alpha}_1=i (-\Delta_1 + i\gamma_1) \alpha_1 - i  \frac{g\omega_1}{\omega_2} \alpha_2  +  F_1 e^{i\phi_1} \;, \\
\dot{\alpha}_2=i (-\Delta_2 + i\gamma_2) \alpha_2 - i  \frac{g\omega_2}{\omega_1} \alpha_1  +  F_2 e^{i\phi_2} \;,
\label{eq4}
 \end{split}
\end{equation}
where a transformation $\phi_i\rightarrow \phi_i+\pi/2$ is considered and $\Delta_i=\omega_i-\omega_d$ is the detuning of the $i$th pendulum from the driving field. 

In steady state, i.e., when $\dot{\alpha}_1=\dot{\alpha}_2\approx 0$, the steady state values of the $\alpha_i=\alpha_{i,ss}$ can be found as 
\begin{equation}
    \begin{split}
        \alpha_{1,ss}=\frac{(\gamma_2 + i \Delta_2)  F_1 e^{i\phi_1}-i g F_2 e^{i \phi_2}} {(\gamma_1 + i \Delta_1)(\gamma_2 + i \Delta_2)+g^2}\;, \\ 
        \alpha_{2,ss}= \frac{(\gamma_1 + i \Delta_1)  F_2 e^{i\phi_2}-i g F_1 e^{i \phi_1}} {(\gamma_2 + i \Delta_2)(\gamma_1 + i \Delta_1)+g^2}\;.
        \label{eq5}
    \end{split}
\end{equation}

\begin{center}
\begin{figure*}
\begin{center}
 \includegraphics[width=.8\textwidth]{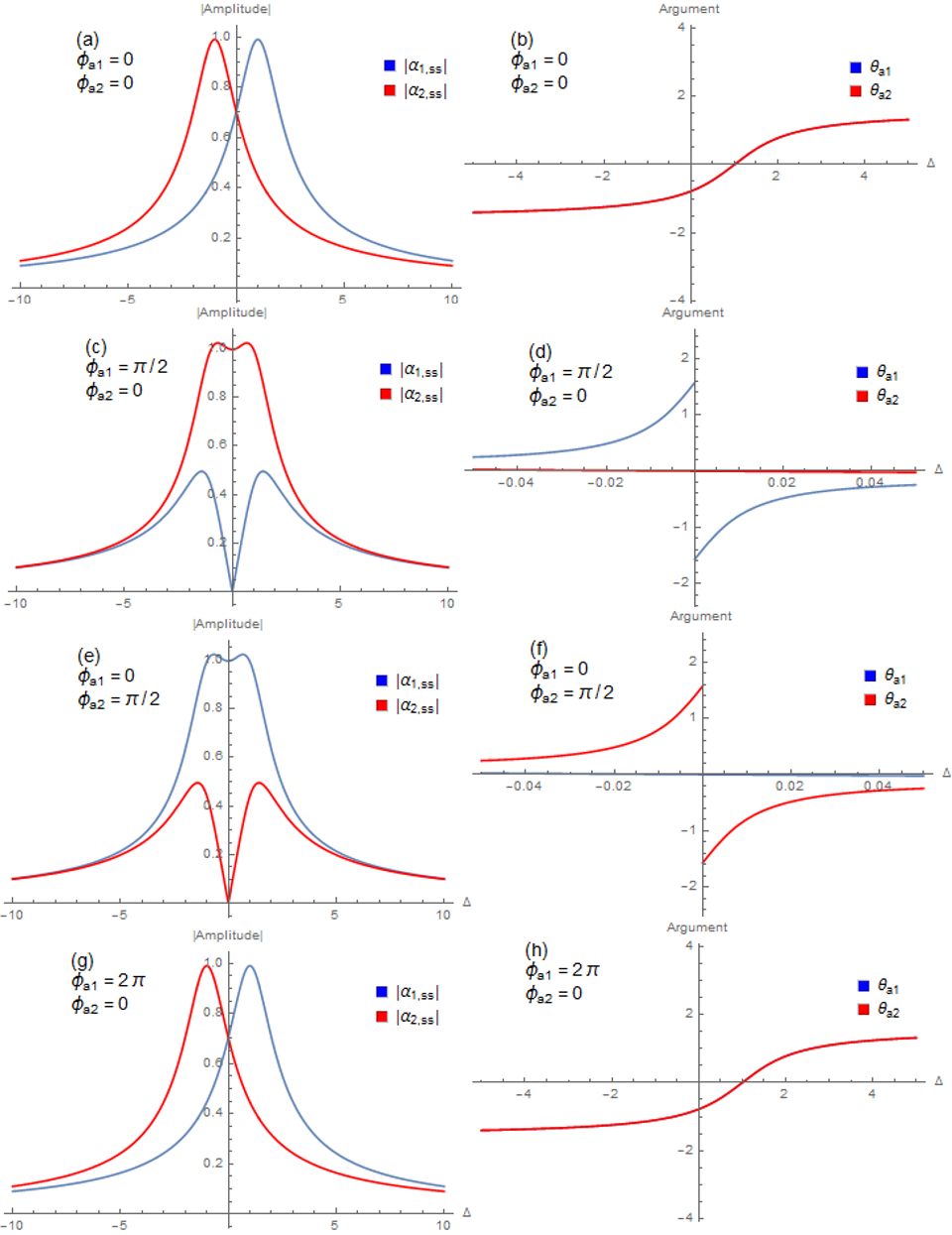}
  \caption{Plots \textbf{(a), (c), (e), (g)} on the left panel represent the variation of magnitudes of the $\alpha_{i,ss}$ with respect to $\Delta$ for various combinations of $\phi_{1}$ and $\phi_{2}$. Plots \textbf{(b), (d), (f), (h)} on the right panel represent the variations of phases of the $\alpha_{i,ss}$ with respect to $\Delta$, corresponding to the same combinations of values of $\phi_{1}$ and $\phi_{2}$. The other parameters chosen are $\gamma_{1}=\gamma_{2}=\gamma$, $F_1 = F_2 = \gamma$, and $g=\gamma$.}
   \label{Fig3}
 \end{center}
 \end{figure*}
\end{center}

We note that the amplitudes are complex. Hence, we plot in Figs. \ref{Fig3} magnitude and phase spectrum of these steady state amplitudes of the oscillators, where we have chosen $\Delta_i=\Delta$. We see that there is a sharp dip in the amplitude of the first oscillator [Fig. \ref{Fig3}(c)], that is associated with a singular phase-change of $\pi$ [Fig. \ref{Fig3}(d)], when the two driving fields have a phase difference $\phi_1-\phi_2=\pi/2$. This phenomenon is called anti-resonance. Such a spectral feature is clearly analogous to non-reciprocity, in which the energy content of the input mode vanishes and that of the output mode becomes maximum. However, for a given phase-difference $\phi_1-\phi_2$, one cannot achieve the reverse situation, namely, vanishing (maximum) energy content of the output (input) mode. This means that the system behaves as an isolator.  We find that The physics of anti-resonance is discussed in \cite{Rao-2019,belbasi2014anti}, where only one of the oscillators is driven by an external harmonic field. 

More importantly, when the phase-difference between the driving fields is changed to $\phi_1-\phi_2=-\pi/2$, we see such anti-resonance behavior in the spectrum of the second oscillator, instead of the first one [Figs. \ref{Fig3}(e) and \ref{Fig3}(f)]. 
Such a reversal of the spectral behavior of two oscillators can also be seen in the context of optomechanical nonreciprocity \cite{Xu-2015,Habraken-2012}, in which such reversal of the directions of the field can be done by changing the phase differences of the two cavity driving fields. We will show explicitly, in the next Section, how these two phenomena are indeed equivalent, not only just analogous. 
We emphasize that by driving only one oscillator, the reversal of the spectral behavior cannot be obtained.


\section{\label{Section3}Equivalence to non-reciprocity and control of isolation frequency by mechanical drive}

In the previous Section, we have outlined how the anti-resonance phenomenon can be analogous to optomechanical nonreciprocity. However the anti-resonance occurs only at a certain frequency $\Delta=0$ [see  Figs. \ref{Fig3}], that depends upon the natural frequency of the mechanical oscillator ($\omega_i$). Analogously, in optomechanical isolators, the isolation occurs only at a certain frequency (which in \cite{Xu-2015} is dependent on natural frequency of the membrane). In such a narrow-band isolator, one is limited with the natural frequency of the membrane itself, that is fixed in a given setup and cannot be manipulated externally. This means that one does not have any dynamical control to make it useful for isolation at other frequencies. In this Section, we will consider a cavity-optomechanical setup to show how such a dynamical control can be achieved. 
\subsection{\label{model}Model}
\begin{center}
\begin{figure*}
\begin{center}
 \includegraphics[width=.8\textwidth]{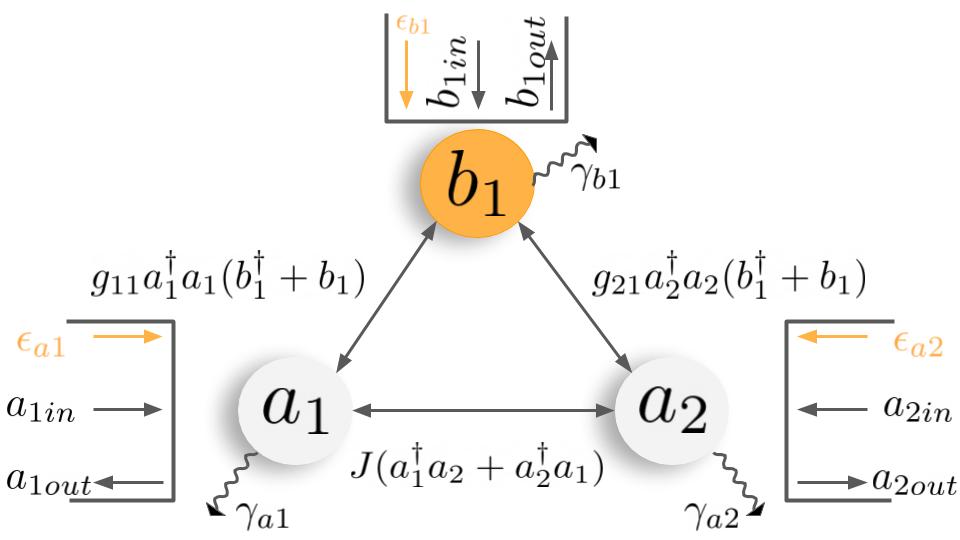}
 \caption{A schematic diagram of the membrane-in-the-middle configuration of a cavity optomechanical system, along with a time-harmonic mechanical drive. All the notations are described in the text.}
 \label{Fig4}
 \end{center}
 \end{figure*}
\end{center}

We start with a ``membrane-in-the-middle" configuration \cite{PhysRevLett.109.013603,PhysRevLett.108.153604,PhysRevLett.109.063601, Wang-2012}, in which the two mirrors of the cavity are kept fixed, while a mechanical oscillator (``the membrane") is suspended inside the cavity [see Fig. \ref{Fig4}]. If this membrane would be fully reflecting on its both sides for the cavity fields, then the fields in the two halves of the cavity could be described by their individual modes $a_1$ and $a_2$. But in our case, in case of the partially reflecting membrane, these modes interact with a coupling constant $J$ and give rise to newer eigenmodes of the entire cavity.  

We further consider that each of the cavity modes is driven by respective external field. In addition, the membrane is also harmonically driven by an external field \cite{Li-2017,Jia-2015,Xu-2015}. The Hamiltonian that governs the dynamics of the entire system can then be written as (in unit of $\hbar = 1)$
\begin{eqnarray}
H &=& \omega_{a1} a_1^\dagger a_1 + \omega_{a2} a_2^\dagger a_2 +  \omega_{b1} b_1^\dagger b_1 + J (a_1^\dagger a_2 + a_2^\dagger a_1)\nonumber\\
&&+ g_{11} a_1^\dagger a_1(b_1^\dagger + b_1) + g_{21} a_2^\dagger a_2(b_1^\dagger + b_1) \nonumber\\
&&+ i \left[\epsilon_{a1}e^{i(\omega_{da1}t+\phi_{a1})}
a_1^\dagger+\epsilon_{a2}e^{i(\omega_{da2}t+\phi_{a2})} a_2^\dagger\right.\nonumber\\
&&\left.+ \epsilon_{b1}e^{i(\omega_{db1}t)} b_1^\dagger - h.c.\right]\;,
    \label{eq6}
    \end{eqnarray}
where $\omega_{a1}$ and $ \omega_{a2}$ are the frequencies of the cavity modes $a_1$ and $a_2$, respectively. $\omega_{b1}$ is the natural frequency of the membrane mode $b_1$, and $g_{11}$ and $g_{21}$ denote the optomechanical coupling constants.The cavity modes are driven by external laser fields with driving frequencies $\omega_{da1}$ and $\omega_{da2}$, with respective amplitudes $\epsilon_{a1}$ and $\epsilon_{a2}$ and phases $\phi_{a1}$ and $\phi_{a2}$. In addition to this, there is a mechanical drive with driving frequency $\omega_{db1}$ and amplitude $\epsilon_{b1}$. 

In the interaction picture with respect to the unperturbed Hamiltonian of the cavity modes and the membrane, the Hamiltonian gets transformed to $H'$, given by $H' = U H U^\dagger - i \hbar U \frac{\partial{U^\dagger}}{\partial{t}}$
where $U = \exp[i(\omega_{da1}t a_1^\dagger a_1 + \omega_{da2}t a_2^\dagger a_2 +\omega_{db1}t b_1^\dagger b_1)]$. Using the Baker-Campbell-Hausdorff formula and the commutation properties of the bosonic operators, we find the following form of the Hamiltonian:
\begin{eqnarray}
H' &=& \Delta_{a1} a_1^\dagger a_1 + \Delta_{a2} a_2^\dagger a_2 +  \Delta_{b1} b_1^\dagger b_1 
+ J (a_1^\dagger a_2 + a_2^\dagger a_1) \nonumber\\
&&+ g_{11} a_1^\dagger a_1(\bar{b_1}^\dagger + \bar{b_1}) +  g_{21} a_2^\dagger a_2(\bar{b_1}^\dagger + \bar{b_1}) 
\nonumber\\ 
&&+i  \left[\epsilon_{a1}e^{i\phi_{a1}} a_1^\dagger +
    \epsilon_{a2}e^{i\phi_{a2}} a_2^\dagger +\epsilon_{b1} b_1^\dagger- h.c.\right]\;,
   \label{eq8}
\end{eqnarray}
where $\Delta_{ai} = \omega_{ai}-\omega_{dai}$ $(i\in 1,2)$ are the detunings of the $i$th cavity mode and $\Delta_{b1}=\omega_{b1}-\omega_{db1}$ is the detuning of the membrane mode, from their respective driving field. Here we have chosen  $\bar{b_1}= b_1 e^{-i \omega_{db1} t}$ 

In this regard, we start with the quantum Langevin's equations for their respective annihilation operators, as given by,
\begin{eqnarray}
\frac{d}{d t} a_1 &=&\left\{-\frac{\gamma_{a_1}}{2}-i\left[\Delta_{a1}+g_{11}\left(b_1+b_1^{\dagger}\right)\right]\right\} a_1-i J a_2 \nonumber\\
&&+\epsilon_{a1} e^{i \phi_{a1}}+\sqrt{\gamma_{a1}} a_{\mathrm{1in}}\;,\\
\label{eq9}
\frac{d}{d t} a_2 &=&\left\{-\frac{\gamma_{a_2}}{2}-i\left[\Delta_{a2}+g_{21}\left(b_1+b_1^{\dagger}\right)\right]\right\} a_2-i J a_1 \nonumber \\
&&+\epsilon_{a2} e^{i \phi_{a2}}+\sqrt{\gamma_{a_2}} a_{\mathrm{2in}}\;,\\
\label{eq10}
\frac{d}{d t} {b_1} &=&\left(-\frac{\gamma_{b_1}}{2}-i \Delta_{b1}\right) \bar{b_1}-i\left(g_{11} a_1^{\dagger} a_1+g_{21} a_2^{\dagger} a_2\right)\nonumber \\
&&+\epsilon_{b1}e^{-i \omega_{db1} t}+\sqrt{\gamma_{b_1}} b_{\mathrm{1in}}\;,
\label{eq11}
\end{eqnarray}
where, $a_{\mathrm{1in}}$, $a_{\mathrm{2in}}$ and $b_{\mathrm{1in}}$ are the input fields with mean values equal to zero, $\gamma_{ai}$ and $\gamma_{b_1}$ are the damping rates of $i$th cavity mode and the membrane mode, respectively.
 
Taking time-average of the above equations and considering that in the steady state, all the time-derivatives vanish, we can obtain the following solutions  for the mean values of the annihilation operators:
\begin{equation}
\begin{aligned}
\langle a_1\rangle &=\alpha=\frac{\left(\frac{\gamma_{a2}}{2}+i \Delta_{a2}^{\prime}\right) \varepsilon_{a1} e^{i \phi_{a1}}-i J \varepsilon_{a2} e^{i \phi_{a2}}}{\left(\frac{\gamma_{a1}}{2}+i \Delta_{a1}^{\prime}\right)\left(\frac{\gamma_{a2}}{2}+i \Delta_{a2}^{\prime}\right)+J^{2}}\;, \\
\langle a_2\rangle &=\beta=\frac{\left(\frac{\gamma_{a1}}{2}+i \Delta_{a1}^{\prime}\right) \varepsilon_{a2} e^{i \phi_{a2}}-i J \varepsilon_{a1} e^{i \phi_{a1}}}{\left(\frac{\gamma_{a1}}{2}+i \Delta_{a1}^{\prime}\right)\left(\frac{\gamma_{a2}}{2}+i \Delta_{a2}^{\prime}\right)+J^{2}}\;, \\
\langle b_1\rangle &=\xi=\frac{-i\left(g_{11}|\alpha|^{2}+g_{21}|\beta|^{2}\right)}{\left(\frac{\gamma_{b1}}{2}+i \Delta_{b1}\right)}\;.
\end{aligned}
\label{eq12}
\end{equation}
Here, $\Delta_{a1}^{\prime}=\Delta_{a1}+g_{11}\left(\xi+\xi^{*}\right)$ and $\Delta_{a2}^{\prime}=\Delta_{a2}+g_{21}\left(\xi+\xi^{*}\right)$ are the effective detunings.  We have assumed $\Delta'_{ai} \approx \Delta_{b1}$ (analogous to that used in the resolved sideband limit) and $\gamma_{ai}, \gamma_{b1}, g_{i1} \ll \Delta_{b1}$ $(i\in 1,2)$.

We can clearly see that the expressions of the steady state amplitude $\alpha$ and $\beta$ of the intracavity fields are similar to those in Eq. (\ref{eq5}). This means that with suitable choice of the coupling $J$ and the driving field amplitudes $\epsilon_{ai}$, one can have anti-resonance in either cavity modes (that corresponds to either $\alpha=0,\beta\neq 0$ or vice versa). The phase difference $\phi_{a1}-\phi_{a2}$ between the driving fields governs in which cavity mode, one can achieve this anti-resonance. In the next Section, we will explicitly show how this is related to the non-reciprocity.

\subsection{Non-reciprocity and control of isolation frequency}
To investigate the nonreciprocity at the steady state, we need to obtain the fluctuation dynamics of all the subsystems involved, namely, two cavity modes and the membrane. We consider that
the driving fields amplitudes are larger than the relevant decay rates. In this limit, we can linearize the Eqs. (\ref{eq9}-\ref{eq11}), by expanding the annihilation operators as a sum of its steady state averages and the fluctuation operators, namely, $a_1 = \alpha + \delta a_1$, $a_2 = \beta + \delta a_2$, $b_1 = \xi + \delta b_1$. Substituting them in the Eqs. (\ref{eq9}-\ref{eq11}), the linearized quantum Langevin equations can be written as

\begin{eqnarray}
\frac{d}{d t} \delta a_1 &=&\left(-\frac{\gamma_{a1}}{2}-i \Delta_{a1}^{\prime}\right) \delta a_1-i G_{11}\left(\delta b_1+\delta b_1^{\dagger}\right) \nonumber\\
&&-i J \delta a_2+\sqrt{\gamma_{a1}} a_{\mathrm{1in}}\;,\\
\label{eq13}
\frac{d}{d t} \delta a_2&=&\left(-\frac{\gamma_{a2}}{2}-i \Delta_{a2}^{\prime}\right) \delta a_2-i G_{21}\left(\delta b_1+\delta b_1^{\dagger}\right) \nonumber\\
&&-i J \delta a_1+\sqrt{\gamma_{a2}} a_{\mathrm{2in}}\;,\\
\label{eq14}
\frac{d}{d t} \delta b_1&=&\left(-\frac{\gamma_{b1}}{2}-i \Delta_{b1}\right) \delta b_1-i\left(G_{11} \delta a_1^{\dagger}+G_{11}^{*} \delta a_1\right) \nonumber\\
&&-i\left(G_{21} \delta a_2^{\dagger}+G_{21}^{*} \delta a_2\right)+\sqrt{\gamma_{b1}} b_{\mathrm{1in}}\;,
\label{eq15}
\end{eqnarray}
where $G_{11}=g_{11} \alpha=\left|G_{11}\right| e^{i \theta_{a1}}$ and  $G_{21}=g_{21} \beta=\left|G_{21}\right| e^{i \theta_{a2}}$. Here $\theta_{ai}$ are the phases of the complex steady state amplitudes $\alpha$ and $\beta$. 
These phases can be easily related to those $\phi_{ai}$ of the driving fields (see in Sec. \ref{appen}).
Note that the above equations contain terms with $\alpha$ and $\beta$. So, in the condition of anti-resonance (when either of them vanishes), the fluctuation dynamics changes.

These equations can be solved in Fourier domain. We first write them in the form of matrix elements as
\begin{equation}
    \dot{V}_i = -M_{ij}V_j + \Gamma_{ij}V_{j,in}
    \label{eq16}
\end{equation}

where $i,j = 1,2,...,2n$ (where $n=3$ is the number of distinct modes in the system). Here the elements of fluctuation and input field vectors are arranged as 
$
V=\left(\delta a_1, \delta a_2, \delta b_1, \delta a_1^{\dagger}, \delta a_2^{\dagger}, \delta b_1^{\dagger}\right)^{T}
$ and $ 
V_{in}=\left(a_{\mathrm{1in}}, a_{\mathrm{2in}}, b_{\mathrm{1in}}, a_{\mathrm{1in}}^{\dagger}, a_{\mathrm{2in}}^{\dagger}, b_{\mathrm{1in}}^{\dagger}\right)^{T}
$, respectively. The matrix $\Gamma$ is given by  
$\operatorname{diag}(\sqrt{\gamma_{a1}}, \sqrt{\gamma_{a2}}, \sqrt{\gamma_{b1}}, \sqrt{\gamma_{a1}}, \sqrt{\gamma_{a2}}, \sqrt{\gamma_{b1}})$.

By introducing the Fourier transform of the operators $\hat{o}$, as 
\begin{eqnarray}
\widetilde{o}(\omega)&=&\frac{1}{\sqrt{2 \pi}} \int_{-\infty}^{+\infty} o(t) e^{i \omega t} d t\;,\\
\label{eq17}
\widetilde{o}(\omega)^{\dagger}&=&\frac{1}{\sqrt{2 \pi}} \int_{-\infty}^{+\infty} o(t)^{\dagger} e^{i \omega t} d t\;,
\label{eq18}
\end{eqnarray}
the matrix equation (\ref{eq16}) can be rewritten in the frequency domain as,
\begin{equation}
\widetilde{V_{i}}(\omega)=(M-i \omega I)_{ij}^{-1} \Gamma_{jk} \widetilde{V}_{\mathrm{k,in}}(\omega)\;,
\label{eq19}
\end{equation}
where $I$ is the identity matrix. 

In the input-output formalism \cite{Gardiner-1985}, the relationship between internal, input and output fields under the first Markov approximation is given as 
\begin{equation}
o_{\mathrm{out}}+o_{\mathrm{in}}=\sqrt{\gamma_{o}} \delta o\;.
\label{eq20}
\end{equation}
where, $o \equiv a_1,a_2,b_1$  and the $\gamma_o$ are the damping rates for the respective $o$. Using this in (\ref{eq19}), and defining $\tilde{V}_{\mathrm{out}}=\left(a_{\mathrm{1,out}}, a_{\mathrm{2,out}}, b_{\mathrm{1,out}}, a_{\mathrm{1,out}}^{\dagger}, a_{\mathrm{2,out}}^{\dagger}, b_{\mathrm{1,out}}^{\dagger}\right)^{T}$, we find that 
\begin{equation}
\tilde{V}_{i, \text{out }}(\omega)=\sum_{l=1}^{2n}U_{i,l}\tilde{V}_{l,\text{in}}(\omega)\;,\;U_{i,l}=[\Gamma_{ij}(M-i \omega I)_{jk}^{-1} \Gamma_{kl}-\delta_{il}]\;.
\label{eq22}
\end{equation}

To obtain the spectrum of scattering probabilities, we now invoke the two-frequency correlation of various elements of the $V$-matrices. Let us start with 
$\tilde{V}_{i,out}^{\dagger}\tilde{V}_{j,out}=\sum_{l,m=1}^{2n}U_{i,l}^{*}U_{j,m}\tilde{V}_{l,in}^{\dagger}\tilde{V}_{m,in}$.
Now, as $\langle \tilde{V}_{i,in}^{\dagger}\tilde{V}_{j,in} \rangle$ are non-zero only for $i=j$, we find that $\langle \tilde{V}_{i,out}^{\dagger}\tilde{V}_{j,out}\rangle=\sum_{l=1}^{2n}U_{i,l}^{*}U_{j,l}\langle \tilde{V}_{l,in}^{\dagger}\tilde{V}_{l,in}\rangle$. 

Defining $S_{i, out}(\omega)=\int  d\omega^{\prime}\left\langle\tilde{V}_{i,out }^{\dagger}\left(\omega^{\prime}\right) \tilde{V}_{i,out }(\omega)\right\rangle$ and 
$S_{i, in}(\omega)=\int  d\omega^{\prime}\left\langle\tilde{V}_{i,in }^{\dagger}\left(\omega^{\prime}\right) \tilde{V}_{i,in }(\omega)\right\rangle$ ($i\in 1, 2, 3$), we now obtain following expression, relating the frequency correlations at the input and output ports:
\begin{equation}
S_{\text {out }}(\omega)=T(\omega) S_{\text {in }}(\omega)+S_{\text {vac }}(\omega)\;,
\label{eq23}
\end{equation}
where $S_{\text{in}}$, $S_{\text{out}}$, and $S_{vac}(\omega)=\left(s_{a1,vac}(\omega), s_{a2,vac}(\omega), s_{b1,vac}(\omega)\right)^{T}$ are the $n$-component column matrices, with $n=3$ (i.e., the number of modes involved). The
$T$ denotes the scattering probability matrix. The element $T_{ij}$ of this matrix represents the probability to scatter from the $i$th mode to the $j$th mode, and is given by
\begin{equation}
    T_{ij}=|U_{i,j}|^2+|U_{i,j+n}|^2\;,
    \label{eq24}
\end{equation}
 for $i, j\in 1,2,3$. Similarly, the elements of $S_{vac}(\omega)$ can be written in terms of the elements of the $U$-matrix, as
\begin{equation}
S_{i,vac}(\omega)=|U_{i,i+n}(\omega)|^{2}+
|U_{i,i+n+1}(\omega)|^{2}+..+|U_{i,i+2n}(\omega)|^{2}\;.
\label{eq25}
\end{equation}

\begin{center}
\begin{figure*}
\begin{center}
 \includegraphics[width=.8\textwidth]{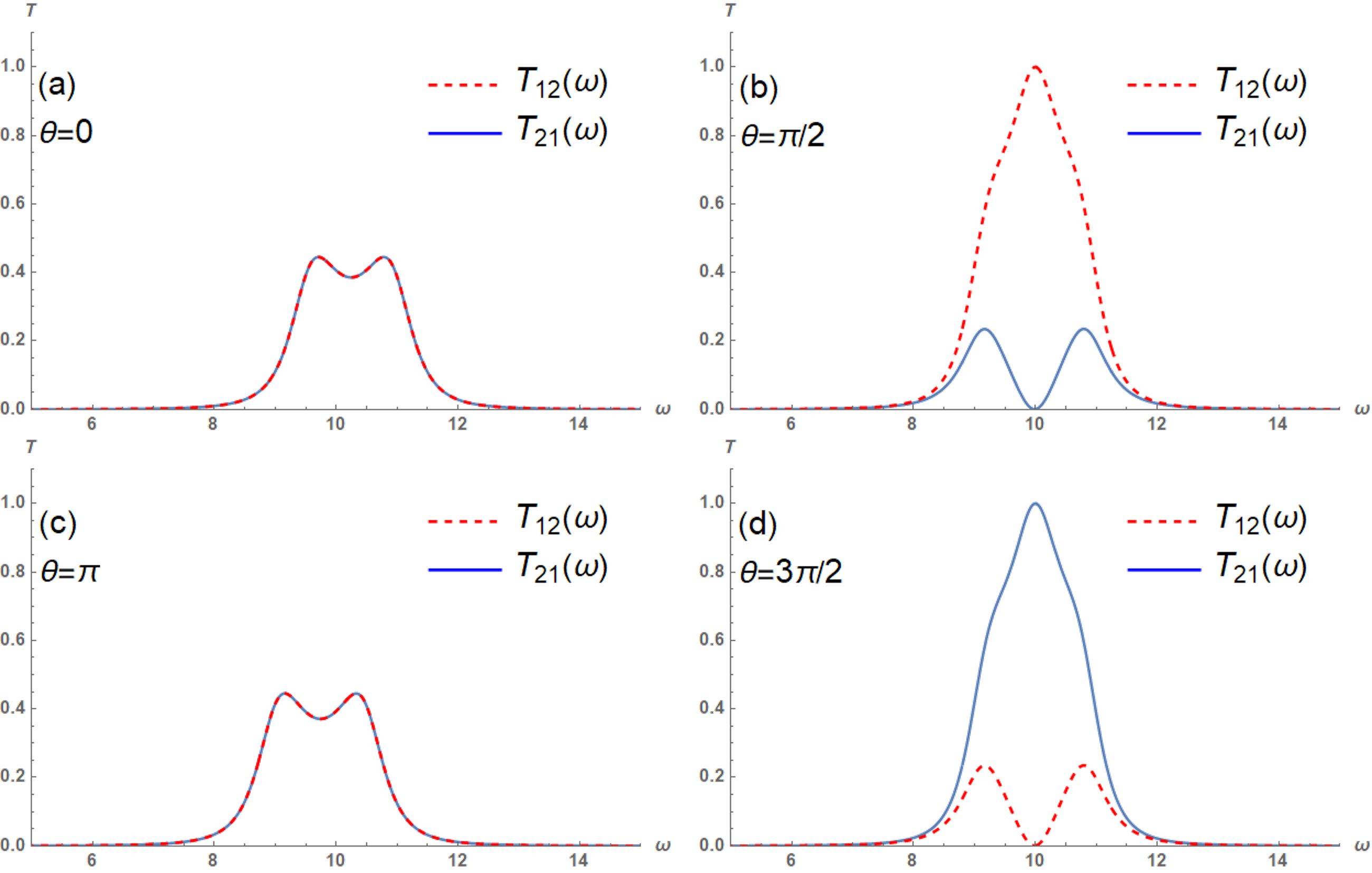} \\
  \caption{Variation of scattering probabilities $T_{12}$ and $T_{21}$ with the normalized input frequency $\omega/\gamma$ for (a) $\theta=0$, (b) $\theta=\pi/2$, (c) $\theta=\pi$, and $\theta=3\pi/2$. The other parameters chosen are  $J=|G_{11}|=|G_{21}|=\gamma_{a1}/2=\gamma_{a2}/2=\gamma_{b1}/2=\gamma/2$ and $\Delta_{b1}=\Delta_{a1}=\Delta_{a2}=10\gamma$.  Clearly, with suitable choice of $\theta$, one can achieve non-reciprocity of in either direction, i.e., either $T_{12}=0$ or $T_{21}=0$.}
  \label{Fig5}
 \end{center}
 \end{figure*}
\end{center}
\begin{center}

\begin{figure*}

\begin{center}
 \includegraphics[width=.4\textwidth]{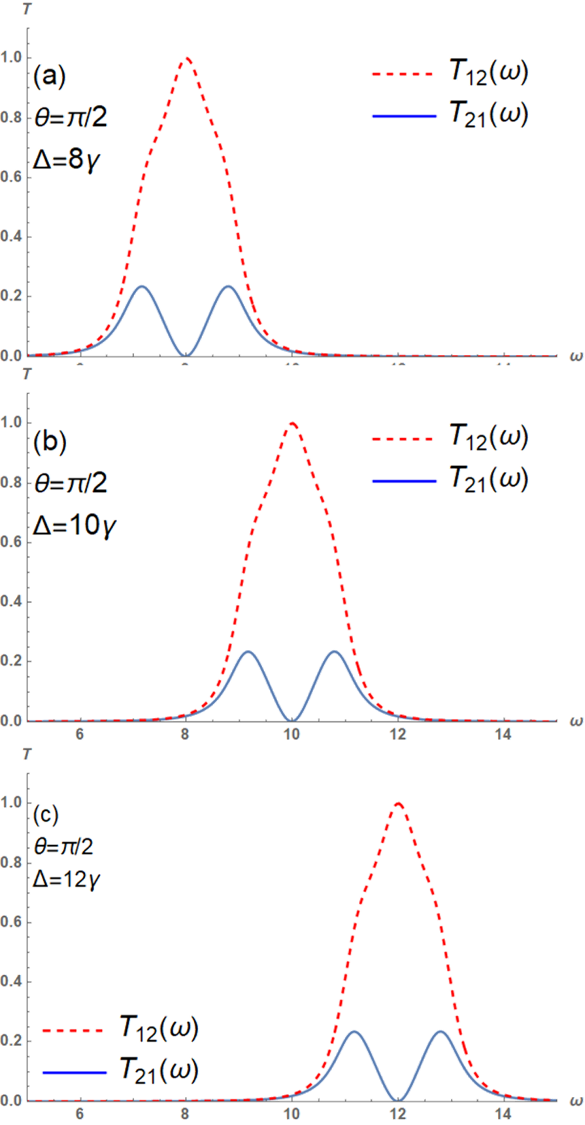} \\
  \caption{Variation of scattering probabilities $T_{12}$ and $T_{21}$ with the normalized input frequency $\omega/\gamma$ for (a) $\Delta=8\gamma$, (b) $\Delta=10\gamma$, and $\Delta = 12\gamma$. 
  We have chosen  the following parameters:  $J=|G_{11}|=|G_{21}|=\gamma_{a1}/2=\gamma_{a2}/2=\gamma_{b1}/2=\gamma/2$,  $\Delta_{b1}=\Delta'_{a1}=\Delta'_{a2}=\Delta$, and $\theta=\pi/2$. It is clearly seen that the isolation frequency  is attained at $\omega=\Delta$ in all these plots.}
  \label{Fig6}
 \end{center}
 \end{figure*}
\end{center}

In Fig. \ref{Fig5}, we display the spectra of the scattering probability $T_{12}$ from the cavity mode $a_1$ to $a_2$ and $T_{21}$ from the cavity mode $a_2$ to $a_1$, with respect to the frequency $\omega$ of the input signal field. We have chosen different values of the phase difference $\theta$ between the effective optomechanical coupling strengths $G_{11}$ and $G_{21}$. We have chosen all the detunings equal as $\Delta_{a1}',\Delta_{a2}',\Delta_{b1} = \Delta=10\gamma$.  It is clearly seen that for $\theta=\pi/2$, the $T_{21}=0$ when the frequency of the field becomes $\omega=\Delta$, while $T_{12}$ becomes unity [see Fig. \ref{Fig5}(b)]. This means that for a certain frequency, one can achieve a complete transmission of the field from the mode $a_1$ to $a_2$, but not vice versa. The scenario can be reversed by changing the phase difference $\theta$ to $3\pi/2$. Generally speaking, optical isolation occurs for $\theta=(2k+1)\pi/2$, for all integers $k$, while for $\theta\equiv \pi/2 (\mathrm{mod}\; 2\pi)$, the transmission vanishes from the mode $a_2$ to $a_1$ and for $\theta\equiv 3\pi/2 (\mathrm{mod} \; 2\pi)$, the reverse situation is achieved. 

The very fact that the membrane mode is driven by an external field (unlike in \cite{Xu-2015}) provides us an additional handle to control the isolation frequency at which one achieves non-reciprocity in either direction. We show in Fig. \ref{Fig6}, how the  scattering probabilities $T_{12}$ and $T_{21}$ vary with $\omega$ for $\theta=\pi/2$, but with varying values of the equal detunings $\Delta$. It is clearly seen that one can achieve optical isolation at $\omega=\Delta$, for any chosen value of $\Delta$. Choosing different values of $\Delta$ corresponds to setting up the frequencies of the driving fields externally. Such a control of isolation frequency cannot be achieved in the models described in \cite{Xu-2015,Habraken-2012,Miri-2017}, where the isolation occurs at the fundamental frequency of the membrane, which cannot be changed in a given setup. The dependence of $\theta$ however remains the same, as displayed in Fig. \ref{Fig5}.  Therefore, it is possible to tune the isolation frequency at any desired value, by suitable choices of phase and frequency of the external fields.

\subsection{\label{appen}Relation between anti-resonance and non-reciprocity}
We have mentioned before that the steady state average values of the field amplitudes in the cavity modes exhibit antiresonance, as the Eqs. (\ref{eq5}) and (\ref{eq12}) are exactly similar. Comparing with the parameters chosen in Fig. \ref{Fig2} for the coupled pendulums, we find that, in case of our model of optomechanical system, one can achieve antiresonance for a coupling constant $J=\gamma_{a1}/2 = \gamma_{a2}/2)$ and $\varepsilon_{a1}=\varepsilon_{a2}$. In fact, we have $\alpha=0$, if $\phi_{a1}-\phi_{a2}=\pi/2$ and $\beta=0$, if $\phi_{a1}-\phi_{a2}=3\pi/2$, all at the zero detuning: $\Delta_{a1}=\Delta_{a2}=0$.

Interestingly, for non-reciprocity in the system, we have chosen the same parameters in Fig. \ref{Fig5} and \ref{Fig6}. Note further that we have used a parameter domain, as $\Delta=  10\gamma$, J $= \gamma/2$, satisfying a large detuning limit $\Delta\gg J, \gamma$. Hence, from Eq. \ref{eq12}, we have, for $\epsilon_{a1}=\epsilon_{a2}=\epsilon$,
\begin{eqnarray}
    \alpha & \approx & \frac{\Delta\epsilon}{D}e^{i(\phi_{a1}+\pi/2-\zeta)}\;,\nonumber\\
    \beta &\approx & \frac{\Delta\epsilon}{D}e^{i(\phi_{a2}+\pi/2-\zeta)}\;.
    \label{eq26}
\end{eqnarray}
Here, the denominator in the expressions of $\alpha$ and $\beta$ [see Eqs. (\ref{eq12})] is written as $D e^{i\zeta}$, and the phase $\pi/2$ appears in the numerator. Clearly, the phase difference between $\alpha$ and $\beta$ is given by $\phi_{a1}-\phi_{a2}$. This is exactly the same as $\theta_{a1}-\theta_{a2}$, as used in the expressions of the effective optomechanical coupling: $G_{a1}=g_{11}\alpha=|G_{a1}|e^{i\theta_{a1}}$ and $G_{a2}=g_{21}\alpha=|G_{a2}|e^{i\theta_{a2}}$, while the optomechanical couplings $g_{11}$ and $g_{21}$ are real quantities. The driving fields therefore create an effective phase-difference in the optomechanical couplings, which in turn, lead to non-reciprocity. This further confirms that the non-reciprocity can be inherently attributed to the anti-resonance, a well-known quantum interference effect of two oscillators.



\section{\label{Section4}Conclusion}
  We have explicitly shown that  an optomechanical isolator works in the principle of anti-resonance. We have shown this by drawing analogy with the anti-resonance of a system of coupled driven oscillators. We have derived the equations for steady state averages of the field amplitudes in the cavity modes and have exploited its similarity with the steady state amplitudes of the oscillators. We further studies the frequency spectrum of the correlations between different modes of the cavity. We find that transmission spectrum indeed exhibits non-reciprocity, in the same condition of anti-resonance. 
In previous studies, it was shown that the isolation frequency, (i.e., the frequency of the input field at which maximum isolation occurs) is limited to a small window which is dependent on inherent system parameters, namely, the fundamental frequency of the membrane \cite{Xu-2015,Habraken-2012,Miri-2017}. This frequency could not be dynamically manipulated. We here have shown that this issue can be resolved by driving the membrane by an external field. The suitable detuning of the mechanical drive allows the control of isolation frequency,  

\section*{acknowledgments}
One of us (C.W.) would like to acknowledge Council of Scientific and  Industrial Research (CSIR), India for financial assistance through CSIR-JRF fellowship.

 are currently exploring non-reciprocity in cavity optomechanical systems \cite{Verhagen-2017}.

\bibliographystyle{plain} 
\bibliography{biblio.bib}

\providecommand{\noopsort}[1]{}\providecommand{\singleletter}[1]{#1}%
\begin{thebibliography}{10}

\bibitem{book:75551}
Malvin Carl~Teich Bahaa E. A.~Saleh.
\newblock {\em Fundamentals of photonics}.
\newblock Wiley Series in Pure and Applied Optics. WILEY, 2ed., wiley edition,
  2007.

\bibitem{belbasi2014anti}
Somayyeh Belbasi, M~Ebrahim~Foulaadvand, and Yong~S Joe.
\newblock Anti-resonance in a one-dimensional chain of driven coupled
  oscillators.
\newblock {\em American Journal of Physics}, 82(1):32--38, 2014.

\bibitem{Caloz-2018}
C.~Caloz, A.~Alù, S.~Tretyakov, D.~Sounas, K.~Achouri, and Z.~DeckLéger.
\newblock Electromagnetic nonreciprocity.
\newblock {\em Physical Review Applied}, 10:047001, 2018.

\bibitem{Fang-2017}
K.~Fang, J.~Luo, A.~Metelmann, Matthew~H. Matheny, F.~Marquardt, A~A. Clerk,
  and O.~Painter.
\newblock Generalized non-reciprocity in an optomechanical circuit via
  synthetic magnetism and reservoir engineering.
\newblock {\em Nature Physics}, 2017.

\bibitem{foulaadvand2010mechanical}
M~Ebrahim Foulaadvand and Davoud Masoumi.
\newblock Mechanical filtering in forced-oscillation of two coupled pendulums.
\newblock {\em arXiv preprint arXiv:1006.2475}, 2010.

\bibitem{Gardiner-1985}
C.~W. Gardiner and M.~J. Collett.
\newblock Input and output in damped quantum systems: Quantum stochastic
  differential equations and the master equation.
\newblock {\em Physical Review A}, 31:3761--3774, 1985.

\bibitem{Alzar-2002}
C.~L. Garrido~Alzar, M.~A.~G. Martinez, and P.~Nussenzveig.
\newblock Classical analog of electromagnetically induced transparency.
\newblock {\em American Journal of Physics}, 70:37, 2002.

\bibitem{Habraken-2012}
S~J~M. Habraken, K.~Stannigel, M~D. Lukin, P.~Zoller, and P~Rabl.
\newblock Continuous mode cooling and phonon routers for phononic quantum
  networks.
\newblock {\em New Journal of Physics}, 14:115004, 2012.

\bibitem{Hemmer-1988}
P.~R. Hemmer and M.~G. Prentiss.
\newblock Coupled-pendulum model of the stimulated resonance raman effect.
\newblock {\em Journal of the Optical Society of America B}, 5:1613, 1988.

\bibitem{Hofstadter-1976}
Douglas Hofstadter.
\newblock Energy levels and wave functions of bloch electrons in rational and
  irrational magnetic fields.
\newblock {\em Physical Review B}, 14:2239--2249, 1976.

\bibitem{Jalas-2013}
D.~Jalas, A.~Petrov, M.~Eich, W.~Freude, S.~Fan, Z.~Yu, R.~Baets, M.~Popović,
  A.~Melloni, J.~D. Joannopoulos, M.~Vanwolleghem, C.~R. Doerr, and H.~Renner.
\newblock What is — and what is not — an optical isolator.
\newblock {\em Nature Photonics}, 7:579--582, 2013.

\bibitem{Jia-2015}
W.~Z. Jia, L.~F. Wei, Yong Li, and Y.~Liu.
\newblock Phase-dependent optical response properties in an optomechanical
  system by coherently driving the mechanical resonator.
\newblock {\em Physical Review A}, 91:043843, 2015.

\bibitem{book:289457}
F.~S.~Crawford Jr.
\newblock {\em Waves}.
\newblock MGH, 1968.

\bibitem{Li-2017}
Y.~Y. Li, Y.and~Huang, X.~Z. Zhang, and L.~Tian.
\newblock Optical directional amplification in a three-mode optomechanical
  system.
\newblock {\em Optics Express}, 25:18907, 2017.

\bibitem{PhysRevLett.109.063601}
Max Ludwig, Amir~H. Safavi-Naeini, Oskar Painter, and Florian Marquardt.
\newblock Enhanced quantum nonlinearities in a two-mode optomechanical system.
\newblock {\em Phys. Rev. Lett.}, 109:063601, Aug 2012.

\bibitem{Metelmann-2015}
A.~Metelmann and A.A. Clerk.
\newblock Nonreciprocal photon transmission and amplification via reservoir
  engineering.
\newblock {\em Physical Review X}, 5:021025, 2015.

\bibitem{Miri-2017}
M.~Miri, F.~Ruesink, E.~Verhagen, and A.~Alù.
\newblock Optical nonreciprocity based on optomechanical coupling.
\newblock {\em Physical Review Applied}, 7:064014, 2017.

\bibitem{Novotny-2010}
L.~Novotny.
\newblock Strong coupling, energy splitting, and level crossings: A classical
  perspective.
\newblock {\em American Journal of Physics}, 78:1199, 2010.

\bibitem{book:2130260}
S.~Rajasekar and M.~A.~F. Sanjuán.
\newblock {\em Nonlinear Resonances, Chap. 14}.
\newblock Springer series in synergetics.; Springer complexity. Springer, 1st
  ed. edition, 2016.

\bibitem{Rao-2019}
J.~Rao, C.~Yu, Y.~Zhao, Y-S. Gui, X.~Fan, D.~Xue, and C-M. Hu.
\newblock Level attraction and level repulsion of magnon coupled with a cavity
  anti-resonance.
\newblock {\em New Journal of Physics}, 2019.

\bibitem{Satpathy-2012}
S.~Satpathy, A.~Roy, and A.~Mohapatra.
\newblock Fano interference in classical oscillators.
\newblock {\em European Journal of Physics}, 33:863--871, 2012.

\bibitem{Souza-2015}
J.~A. Souza, L.~Cabral, R.~R. Oliveira, and C.~J. Villas-Boas.
\newblock Electromagnetically-induced-transparency-related phenomena and their
  mechanical analogs.
\newblock {\em Physical Review A}, 92:023818, 2015.

\bibitem{PhysRevLett.109.013603}
K.~Stannigel, P.~Komar, S.~J.~M. Habraken, S.~D. Bennett, M.~D. Lukin,
  P.~Zoller, and P.~Rabl.
\newblock Optomechanical quantum information processing with photons and
  phonons.
\newblock {\em Phys. Rev. Lett.}, 109:013603, Jul 2012.

\bibitem{PhysRevLett.108.153604}
Lin Tian.
\newblock Adiabatic state conversion and pulse transmission in optomechanical
  systems.
\newblock {\em Phys. Rev. Lett.}, 108:153604, Apr 2012.

\bibitem{Verhagen-2017}
Andrea Verhagen, Ewold;~Alù.
\newblock Optomechanical nonreciprocity.
\newblock {\em Nature Physics}, 13:922--924, 2017.

\bibitem{Wang-2012}
Aashish~A Wang, Ying-Dan;~Clerk.
\newblock Using dark modes for high-fidelity optomechanical quantum state
  transfer.
\newblock {\em New Journal of Physics}, 14:105010, 2012.

\bibitem{Xu-2015}
Yong Xu, Xun-Wei;~Li.
\newblock Optical nonreciprocity and optomechanical circulator in three-mode
  optomechanical systems.
\newblock {\em Physical Review A}, 91:053854, 2015.

\end{thebibliography}
\end{document}